\documentclass[pre,twocolumn,showpacs,amsmath,amssymb,cite]{revtex4-1}

\usepackage{graphicx}
\usepackage{epsfig}
\usepackage{dcolumn}
\usepackage{bm}

\begin{document}
\title{A Structural Model for Fluctuations in Financial Markets}

\author{Kartik Anand$^1$, Jonathan Khedair$^2$, and Reimer K\"uhn$^2$}
\affiliation{$^1$Deutsche Bundesbank, Wilhelm-Epstein-Strasse 14, 60431 Frankfurt am Main, Germany\\
$^2$Department of Mathematics, King's College London, Strand, London WC2R 2LS, UK}

\date{\today}

\begin{abstract}
In this paper$^*$ we provide a comprehensive analysis of a structural model for the dynamics of prices of assets traded in a market originally proposed in \cite{KuNeu08}. The model takes the form of an interacting generalization of the geometric Brownian motion model. It is formally equivalent to a model describing the stochastic dynamics of a system of analogue neurons, which is expected to exhibit glassy properties and thus many meta-stable states in a large portion of its parameter space. We perform a generating functional analysis, introducing a slow driving of the dynamics to mimic the effect of slowly varying macro-economic conditions. Distributions of asset returns over various time separations are evaluated analytically and are found to be fat-tailed in a manner broadly in line with empirical observations. Our model also allows to identify collective, interaction mediated properties of pricing distributions and it predicts pricing distributions which are significantly broader than their non-interacting counterparts, if interactions between prices in the model contain a ferro-magnetic bias. Using simulations, we are able to substantiate one of the main hypotheses underlying the original modelling, viz. that the phenomenon of volatility clustering can be rationalised in terms of an interplay between the dynamics within meta-stable states and the dynamics of occasional transitions between them.\\[2.5mm]
$^*$The opinions expressed in this paper are those of the authors and do not necessarily reflect the views of the Deutsche Bundesbank, the Eurosystem, or their staff.
\end{abstract}



\maketitle
\section{Introduction}
Predicting and measuring the risk that the value of an investment portfolio will depreciate is a mainstay of financial mathematics. Integral to the success of these endeavours is identifying the various market risk (MR) factors and developing models for their evolution. These MRs include, amongst others, fluctuations in stock indices, changes in interest rates, foreign exchange parities or commodity (e.g., gold, oil, etc.) prices.

In recognition of the importance of MRs, the Basel Committee for Banking Supervision (BCBS) stipulates that banks must explicitly reserve a portion of equity capital against MR. The Basel III accord \cite{Basel16} defines MR as the ``the risk of losses arising from movements in market prices". It proposes two approaches to measure MRs, a so-called Standardized Approach, and an Internal Models Approach. 

Within the standardized approach the market risk of a set of trading positions is defined by their exposure to a standardized set of risk factors and measured in terms of sensitivities of the market values of these positions to movements of the risk factors. To determine the capital to be held against MRs, the risks of various positions held by a bank are aggregated, using prescribed risk weights and prescribed correlations between risk factors.  

Under the Internal Models Approach, banks are permitted to design their own measurement model, which must adhere to strict guidelines. Guidelines cover a host of qualitative and quantitative standards. At a minimum, internal models must encompass the positions covered by the standard model, be regularly back-tested against historical market data to demonstrate their adequacy and accuracy, and be supplemented by a regular and rigorous regime of stress testing. 

The collapse of the US based hedge fund Long-Term Capital Management (LTCM) in 2000, following the East-Asian financial crisis of mid 1997 \cite{AsianCrisis1997} is a poignant example of adverse MRs spreading across wide geographic regions. The crisis was instigated by a devaluation of Thailand's currency, the Thai baht. This move sent shock waves through the economies of East-Asian countries, thereby triggering recessions. The economic downturns led to a sharp decline in the demand and price of oil. Russia, as a major oil producing country, was adversely hit and defaulted on its' public debt. The culmination of all these interlinked shocks resulted in huge losses for LTCM and, in early 2000, its liquidation. It can fairly be argued that chains of events such as this call for a {\em more interconnected\/} approach to market risks than stipulated by the BCBS, even today.

The credit crunch of 2007-09 provides further evidence to support the idea of interactions between risky market positions. While financial derivatives and globalization, which allow for greater portfolio diversification, may have helped mitigate MRs, adverse feedback loops between financial markets and the real economy may be responsible for propagating asset price shocks across borders and even commodity classes. A more thorough understanding of the dynamics of asset prices would thus be welcome.

A model of MRs should reproduce a set of stylized facts found from empirical analysis of time-series data of returns on investment \cite{Mandelbrot63,Fama65, Gopi+99, Gopi+99b, BouPott06}, i.e., (i) return distributions are ``fat-tailed", (ii) the variance of returns is time dependent, and (iii) there are long-range  correlations between variance of returns in time, a phenomenon referred to as ``volatility clustering". 

Over the years a variety of {\em descriptive\/} models have been developed for the returns in financial time-series. These models do not attempt to advance theories of the mechanisms underlying price processes, but concentrate on capturing their statistics. Examples include auto-regressive \cite{Bollersev86,Bollersev87} and stochastic volatility models \cite{Ruiz94,Broto04}. Other models assume that the statistics for return-increments follow symmetric \cite{Mandelbrot63} or asymmetric \cite{Tucker92} stable Paretian distributions.

In an alternative {\em structural\/} model approach one attempts to model the {\em mechanisms\/} behind market dynamics. One possible formulation in this regard, is to consider the collective results of actions performed by agents operating in the market. Models of this type include the minority game \cite{ChalletZhang97, CoolenMG05} and percolation models \cite{StaufferSorn99, ContBou00}.

In \cite{KuNeu08} the authors take an intermediate approach and propose an interacting variant of the geometric Brownian motion model (henceforth referred to as the iGBM) as a structural model of  asset price dynamics. They suggest that the structure of such a model should follow from very generic considerations concerning market mechanisms, arguing in particular that the dynamical evolution of a market, when reduced to a description in terms of asset price dynamics, quite generally {\em must\/} exhibit ``\dots interaction[s] between prices, which may be thought of as arising effectively through the collection of agents, each acting on the basis of his or her own, more or less rational perception of the underlying economy and market mechanisms \cite{KuNeu08}". 

Using simulations, the authors go on to demonstrate that such a model is capable of reproducing the main stylized facts for asset returns. Moreover, analytic investigations revealed that, in a significant portion of the space of model parameters, the system is ``glassy" and is therefore expected to exhibit a large number of meta-stable states. The authors argue that it is above all the interplay between dynamics {\em within\/} meta-stable states and {\em occasional transitions between them\/} --- whether spontaneous or induced  by external stimuli --- which accounts for the phenomenon of volatility clustering.

The purpose of the present paper is to provide a more thorough analysis of the iGBM proposed in \cite{KuNeu08}. Specifically, we perform a generating functional analysis of the model in the limit of large system size. We introduce a slow driving of the dynamics to mimic the effect of slowly varying macro-economic conditions, and investigate statistical properties of asset returns by recourse to a separation of time-scales argument, assuming that the system equilibrates at given values of the slow variable describing macro-economic conditions. 
This analysis allows one to compute distributions of asset returns on various time-scales, and it also exposes interesting collective effects on the pricing of assets, which are driven by a combination of the macro-economic driving and the effects of imitation as encoded in the couplings. 

The remainder of this paper is organized as follows. In Sec. \ref{sec: Model Definition} we introduce the model. Sec.~\ref{sec:modsol} provides a solution based on a Generating Functional Analysis (GFA), with technical details of that analysis relegated to an appendix. Phase diagrams are provided in Sec \ref{sec: Results} along with results of return distributions predicted by the model at various time-scales. By looking at a variant of the model which has meta-stable states of a {\em known structure\/} embedded in its couplings, we are in a position to elucidate  in some detail the relation between meta-stable states with a dynamics switching between them at longer time scales on the one hand side, and volatility clustering on the other hand side. Finally, in Sec \ref{sec: Summary} we provide a summary and a concluding discussion.

\section{\label{sec: Model Definition} Model Definitions}

In this section we describe the model for asset price dynamics as introduced in \cite{KuNeu08}. One considers a system consisting of $N$ assets, labelled $i\,=\,1,\,\ldots,\,N$. To each asset $i$, one associates a time dependent price $S_i(t)\,>\,0$. The geometric Brownian motion model postulates that the relative change of the price performs a random walk  captured by the Langevin equation
\begin{equation}
\frac{1}{S_i(t)}\frac{{\rm d}}{{\rm d}t}S_i(t) = \mu_i + \sigma_i\,\xi_i(t)\,,
\end{equation}
where $\xi_i(t)\in\mathbb{R}$ denotes a Gaussian white noise with zero mean and unit variance. The factor $\sigma_i\geq0$ measures the strength of the Gaussian fluctuations, and $\mu_i\geq 0$ characterizes the growth rate.  Defining a normalized log-price  $u_i(t)\,=\,\log[S_i(t)/S_{i0}]$, in which $S_{i_0}$ is a reference price (needed to non-dimensionalize the argument of the logarithm) we obtain a new stochastic differential equation
\begin{equation}
\label{eq: after ito}
\frac{{\rm d}}{{\rm d}t}u_i(t) = I_i + \sigma_i\,\xi_i(t)\,,
\end{equation}
where $I_i = \mu_i - \sigma_i^2/2$ by an application of Ito's lemma. The iGBM model, at the level of the log-prices $u_i(t)$,  is now constructed by  introducing three extra terms into Eq.~(\ref{eq: after ito}), to give us
\begin{eqnarray}
\label{eq: langevin}
\frac{{\rm d}}{{\rm d}t}u_i(t)\,&=&-\kappa_i\,u_i(t)\,+\,\sum_{j=1}^NJ_{ij}\,g(u_j(t))+\,\sigma_0\,u_0(t)\nonumber\\
& & +\,I_i\, +\,\sigma_i\,\xi_i(t)\, .
\end{eqnarray}
The first additional term describes what might be thought of as an effect of fundamentalist traders in the market, creating a mean reversion effect with reversion coefficients $\kappa_i\,>\,0$.  The natural interpretation of the normalizing factors $S_{i0}$ introduced above would in that case be that of `rational prices' of traded assets. The second additional term in Eq.~(\ref{eq: langevin}) describes the interactions between log-prices of mutually dependent assets. We choose the interaction to be most sensitive in the vicinity of the rational prices, by taking $g(u)$ to be a nonlinear, sigmoid function, describing the feedback mechanism. Possible choices for this function include, the error function or hyperbolic-tangent. The strength of this influence is given by $J_{ij}\,\in\,\mathbb{R}$. The sign of $J_{ij}$ depends on the nature of mutual interactions. If, for example, assets $i$ and $j$ refer to firms with mutually beneficial economic relations, one would have $J_{ij}\,>\,0$.  Conversely if they refer to two competing firms, a negative shock on asset $j$, i.e., $g(u_j)\,<\,0$, may positively affect asset $i$, implying $J_{ij}\,<\,0$.  Finally, the $u_0$ term is introduced to act as a global risk component mimicking slowly evolving economic conditions affecting prices of all traded assets. In the present paper we will model the $u_0$ term as a (slow) Ornstein-Uhlenbeck process,
\begin{equation}
\dot{u}_0(t)=-\gamma u_0(t)+\sqrt{2\gamma}\,\xi_0(t),
\label{OU}
\end{equation} 
where we take $\gamma\ll1$ such that the above process becomes considerably slower than the microscopic asset dynamics described in Eq.~(\ref{eq: langevin}).

For symmetrically coupled networks, the combined effect of mean reversion and the sigmoid nature of the feedback function is known to render such systems stable at long times \cite{Hopfield84}.

As noted in \cite{KuNeu08}, the model is formally equivalent to a model describing the stochastic evolution of a system of graded response neurons \cite{Hopfield84}, with the $u_i(t)$ playing the role of a post-synaptic potentials, with $g(u)$ describing the neuronal input-output relation, the $\kappa_i$ representing  trans-membrane conductances and the $J_{ij}$ the synaptic efficacies. The $I_i$ finally represent external (sensory) inputs, and the $u_0$ term --- not typically included in the original neural modelling \cite{Hopfield84} --- could describe the effects of neuro-modulators.

A lot is know about systems of this type \cite{Hopfield84, SomCriSom88, KuBovH91, FukShii90, Waugh+90, ShiiFuk92, Molgedey+92, KuehnBoes93, KuNeu08}. For the purpose of the present paper, the most important feature is that iGBM type models as described by Eqs.~(\ref{eq: langevin}) and (\ref{OU}) are --- in a large part of their parameter space --- expected to exhibit glassy phases \cite{KuBovH91,ShiiFuk92, KuehnBoes93, KuNeu08} characterized by the existence of a very large number of long-lived meta-stable states \cite{FukShii90, Waugh+90}. The hypothesis investigated in \cite{KuNeu08} was that it would be the interplay between {\em dynamics within meta-stable states\/} and the dynamics of (occasional) {\em transitions between them}, which could be held responsible for the intermittent dynamics of financial markets.

For the purposes of the present analytic study we will keep a synthetic stochastic setting by taking the $J_{ij}$ to
be of the form
\begin{equation}
J_{ij}\,=\,c_{ij}\,\tilde{J}_{ij}\ ,
\label{Jij}
\end{equation}
where the $c_{ij}\,\in\,\{0,1\}$ are connectivity coefficients describing whether or not an interaction between the prices of assets $i$ and $j$ exists, and the $\tilde{J}_{ij}\,\in\,\mathbb{R}$ describe the strengths of the interactions. We assume that $C=(c_{ij})$ is the adjacency matrix of an Erd\H{os}-R\'enyi random graph of mean degree $c$, but will specialize to the regime of sparse yet large connectivity by taking the limits $N\to\infty$ and $c\to\infty$, with $c/N\to0$. 

The $\tilde{J}_{ij}$ are taken to be quenched random quantities with mean and variance scaling with the mean connectivity $c$ to ensure the existence of the large system, i.e. we put
\begin{equation}
\tilde{J}_{ij}\,=\,\frac{J_0}{c}\,+\,\frac{J}{\sqrt{c}}x_{ij}\,,
\label{jtilde}
\end{equation}
in which the $x_{ij}$ are zero mean and unit variance random variables chosen to be independent in pairs with $\overline{x^{\phantom x}_{ij}x_{ji}}\,=\alpha$. The parameter $\alpha\,\in\,[-1,1]$ thus describes the degree of correlations between $\tilde{J}_{ij}$ and $\tilde{J}_{ji}$, with fully symmetric interactions given by $\alpha\,=\,1$. It turns out that the collective properties of such a dilute system in the large mean connectivity limit are actually indistinguishable from those of a fully connected system.

\section{Model Solution}
\label{sec:modsol}

In this section we investigate the dynamics and stationary states for the model introduced in Sec. \ref{sec: Model Definition}. An analysis of the collective properties of the system in the noiseless limit $\sigma_0=\sigma_i=0$ was presented in \cite{KuNeu08}, and used to identify parameter ranges, viz. the regions of small $\kappa_i$ and sufficiently large $J$, where the system would exhibit a large number of meta-stable states. We will demonstrate in Sec.~\ref{sec:MSVolC} below that our microscopic model described by Eq.~(\ref{eq: langevin}) does indeed produce intermittent dynamics in this parameter range.

An exact and formal treatment of the {\em dynamics\/} is possible using a generating functional analysis (GFA) \cite{Dominicis78, SomZipp82, HatchettCoo04, CastCav05}, to which we now turn. For systems of the type considered here the analysis closely follows \cite{HatchettCoo04}.

\subsection{\label{sec: Model Solution GFA}Generating Functional Analysis}
In what follows we present a solution of the model based on the generating functional formalism, which provides tools for the evaluation of correlation and response functions in terms of a characteristic functional of path probabilities. Performing the average over bond disorder in the sum over dynamical trajectories, details of which are found in Appendix \ref{apx:GFA}, one obtains a family of continuous-time effective single site processes, each parameterized by a specific  combination of single node parameters $\vartheta\equiv(I,\kappa,\sigma)$,
\begin{eqnarray}
\nonumber \dot{u}_{\vartheta}(t) &=& -\,\kappa\,u_{\vartheta}(t)\,+I+\,J_0\,m(t)\,+ \sigma_0u_0(t)\\
	& & + \alpha\,J^2\int_{0}^{t}{\rm d}s\, G(t,\,s)\,n_{\vartheta}(s)\,+\,\phi(t)\,
\label{ss-proc} 
\end{eqnarray}
where $n_\vartheta(s)\,=\,g(u_\vartheta(s))$. The noise  $\phi(t)$ in Eq.~(\ref{ss-proc}) is coloured Gaussian noise, with
\begin{eqnarray}
\langle\,\phi(t)\,\rangle&=& 0\,,
\label{phiav}\\
\langle\,\phi(t)\,\phi(s)\,\rangle &=&  \sigma^2\,\delta(t\,-\,s) + J^2q(t,s)\,.
\label{phicorr}
\end{eqnarray}
The order parameters $m(t)$ and $q(t,s)$ appearing in the equation of motion (\ref{ss-proc}) and in the specification
(\ref{phiav}), (\ref{phicorr}) of the noise statistics must be determined self-consistently to satisfy
\begin{eqnarray}
m(t) &=& \left\langle\left\langle n_{\vartheta}(t) \right\rangle\right\rangle_{\vartheta}\,,\\
q(t,s) &=& \left\langle\left\langle n_{\vartheta}(t)\,n_{\vartheta}(s)\right\rangle\right\rangle_{\vartheta}\,,\\
G(t,s) &=& \left\langle \frac{\delta \langle n_{\vartheta}(t)\rangle}{\delta \phi(s)}\right\rangle_{\vartheta}\,.
\end{eqnarray}
Here, the inner average $\langle \ldots \rangle$ refers to an average over coloured-noise $\phi$ for a given member of the single-site ensemble. The outer average $\langle \ldots \rangle_{\vartheta}$ refers to an average over the ensemble as characterised by the $\vartheta$ distribution. Further details for this calculation are provided in Appendix \ref{apx:GFA}. It should also be noted that this formalism is exact in the $N\rightarrow \infty$ limit.

We highlight the following: (i) there is a dependence of the single site dynamics on the overall `magnetization' $m(t)$;  (ii) for {\em any\/} degree of symmetry of the interactions, $\alpha \neq 0$, the effective single node dynamics is non-Markovian, with memory given by the response function $G(t,s)$; (iii) the noise appearing in the single site dynamics is coloured, with correlations determined by the average temporal correlation $q(t,s)$ of single nodes as described by Eq.~(\ref{phicorr}).  

We have thus reduced our system of of equations describing the dynamics of prices of $N$ interacting assets to an ensemble of dynamical evolution equations self-consistently coupled via a set of order parameters, which becomes exact in the thermodynamic limit. As is usually the case with the GFA, the resulting effective equation of motion is highly non trivial and usually relies on sensible assumptions to be analysed further.

\subsection{Separation of Time Scales --- Quasi-Stationary Regime}

For sufficiently small values of $\gamma$ in Eq.~(\ref{OU}) one expects a separation of dynamical time scales to occur, entailing that the fast $u_{\vartheta}(t)$ processes become statistically stationary on timescales on which the slow $u_0(t)$ process can be treated as non-varying. In what follows we shall thus analyse the $u_{\vartheta}(t)$ dynamics under the assumption that it is stationary at a given value $u_0$ of the slow process.

To assist our analysis further, we approximate Eq.~(\ref{ss-proc}) by neglecting fluctuations in the memory term,  rewriting it as
\begin{eqnarray}
\nonumber \dot{u}_{\vartheta}(t) &=& -\,\kappa\,u_{\vartheta}(t)\,+I+\,J_0\,m(t)\,+ \sigma_0u_0(t)\\
 & &+ \alpha\,J^2\int_{0}^{t}{\rm d}s\, G(t,\,s)\,\langle n_{\vartheta}(s)\rangle\,+\,\phi(t)\,
\label{ss-proc2}
\end{eqnarray}
in which, averages over the effective single process dynamics at given $\vartheta$ appear in the retarded interaction. We are thereby discarding one source of noise in the dynamics, and so are likely to overestimate values of macroscopic order parameters. The important qualitative aspects of the collective properties of the system are, however, expected to remain intact as we shall verify through simulations later on.

Assuming stationarity and time translational invariance for a given $u_0$, we introduce the integrated response
\begin{equation}
\chi\,=\int_{0}^t {\rm d}s\,G(t,s)\,,
\end{equation}
and assume it to remain finite. This allows us to rewrite the effective dynamics in the stationary regime as 
\begin{eqnarray}
\nonumber \dot{u}_{\vartheta}(t) &=& -\,\kappa\,u_{\vartheta}(t)\,+I+\,J_0\,m\,+ \sigma_0u_0\\
	 & &+ \alpha\,J^2\chi m_{\vartheta}\,+\,\phi(t)\,
\label{ss-proc3}
\end{eqnarray}
where $m_{\vartheta}(s)=\langle n_{\vartheta}(s)\rangle$ can be regarded as independent of $s$ for $s$ sufficiently close to $t$ for the response function to be non-negligible.

Anticipating that the correlation $q(t,s)$ might develop a time-persistent value $q$, 
\begin{equation}
q(t,s) \to q\ , \qquad \mbox{as} \qquad |t-s|\to \infty\,,
\end{equation}
we decompose the coloured noise $\phi$ into a static (frozen) and an independent time-varying component
\begin{equation}
\phi (t) = J\sqrt{q} z + \eta(t)\,,
\end{equation}
in which $z\sim {\cal N}(0,1)$, and the statistics of the time varying part of the noise is given by
\begin{equation}
\langle \eta(t)\rangle = 0 ~~,~~ \langle \eta(t)\,\eta(s) \rangle = \sigma^2 \delta(t-s) + J^2 {\cal C}(t,s)\,,
\label{eta.stats}
\end{equation}
with
\begin{equation}
{\cal C}(t,s) = q(t,s)-q \to 0\ , \qquad \mbox{as} \qquad |t-s|\to \infty.
\label{Cts}
\end{equation}

The effective single-process dynamics within the stationary regime can then be rewritten in a more suggestive form as
\begin{equation}
\dot{u}_\vartheta(t)=-\kappa\left(u_\vartheta(t)-\overline u_\vartheta\right) +\eta(t)\ ,
\label{ss-proc*}
\end{equation}
in which we have introduced the (long-term) average
\begin{equation}
\overline u_{\vartheta} = \frac{1}{\kappa}\left[I + J_0 m +J\sqrt{q} z+ \alpha\, J^2 \,\chi \,
m_\vartheta +\sigma_0u_0 \right]\,.
\label{ubar}
\end{equation}
We note that $m_\vartheta=\langle g(u_\vartheta(t))\rangle$, where the average is over the stationary $u_\vartheta$ distribution that has $\overline u_{\vartheta}$ as its long-term mean; thus Eq.~(\ref{ubar}) is a self-consistency equation for the value of this long-term mean. Note also that $\vartheta$ now includes $z$, i.e., $\vartheta=(I,\,\kappa,\,\sigma,\,z)$. 
The solution to Eq.~(\ref{ss-proc*}) is now easily written down as
\begin{equation}
u_\vartheta(t) = \overline u_\vartheta + \Big(u_\vartheta(0)-\overline u_\vartheta\Big)\,{\rm e}^{-\kappa t} + \int_0^t
{\rm e}^{-\kappa (t-s)}\eta(s)\,{\rm d}s\,,
\label{eq:eqbm sln}
\end{equation}
implying that $u_\vartheta(t)$ is a Gaussian process with expectation 
\begin{equation}
 \langle u_\vartheta(t)\rangle=\overline u_\vartheta + \Big(u_\vartheta(0)-\overline u_\vartheta\Big)\,{\rm e}^{-\kappa t}.
\end{equation}
For the auto-covariance 
$$
{\cal C}_{u_\vartheta}(t,t^\prime) = \Big\langle \big(u_\vartheta(t)-\langle u_\vartheta(t)\big) \big(u_\vartheta(t') \langle u_\vartheta(t')\big)\rangle\Big\rangle
$$ 
of the $u_\vartheta(t)$ in the large time limit we get a stationary law depending only on time differences, ${\cal C}_{u_\vartheta}(t,t^\prime)={\cal C}_{u_\vartheta}(t-t^\prime)$ with
\begin{equation}
{\cal C}_{u_\vartheta}(t-t^\prime)= \,\frac{1}{2\kappa}\left(\sigma^2 {\rm e}^{-\kappa |t-t'|} + J^2 \hat {\cal C}(0) \right)\,,
\end{equation}
in which $\hat {\cal C}(0)$ is the zero-frequency limit of the Fourier transform $\hat {\cal C}(\omega) = \int_{-\infty}^\infty {\rm d} s\, {\rm e}^{-i\,\omega\,s} {\cal C}(s)$. It is useful to specifically record the equal time limit of ${\cal C}_{u_\vartheta}$,
\begin{equation}
{\cal C}_{u_\vartheta}(0)=\frac{1}{2\kappa}\left(\sigma^2\,+\,J^2\hat{\cal C}(0)\right)\equiv\sigma^2_{u_\vartheta}.
\label{uvar}
\end{equation}

\subsection{Self-Consistency Equations for the Quasi-Stationary Regime}
With full knowledge of the statistics of the $u_\vartheta(t)$ we can reformulate the self-consistent equations for the order parameters describing the stationary regime. They are (i) the stationary magnetization $m$, (ii) the time persistent part $q$ of the node auto-correlations, (iii) the integrated response $\chi$, and (iv) the zero-frequency limit $\hat{\cal C}(0)$ of the Fourier transform of the (non time-persistent) part ${\cal C}(\tau)$ of the node auto-correlations in the stationary regime. To compute the latter, we also have to evaluate the average node auto-correlations $q(\tau)$. 

To formulate the self-consistency equation for $m = \langle m_\vartheta\rangle_\vartheta$, we recall that $m_\vartheta=\langle g(u_\vartheta(t))\rangle$ where the average is over the stationary $u_\vartheta$ distribution, and hence can be rewritten  as
\begin{equation}
m_\vartheta=\langle g(\overline u_\vartheta + \sigma_{u_\vartheta} x)\rangle_x
\end{equation}
with $\sigma_{u_\vartheta}$ defined in Eq.~(\ref{uvar}), and $\langle\dots\rangle_x$ denoting an average over a ${\cal N}(0,1)$ Gaussian $x$. By definition, an average over the distribution of the set of parameters $\vartheta$ then gives $m=\langle m_\vartheta\rangle_\vartheta$.
Following the same logic for the two-point function $q(\tau)$, we obtain the following full set of self-consistency equations for the order parameters
\begin{eqnarray}
m &=& \Big\langle\Big\langle g(\overline u_\vartheta + \sigma_{u_\vartheta} x)\Big\rangle_x\Big\rangle_{\vartheta}\,,
\label{fpe.m}\\
q(\tau) &=& \Big\langle\Big\langle g\big(\overline u_\vartheta +\sigma_{u_\vartheta} x\big)\ 
g\big(\overline u_\vartheta +  \sigma_{u_\vartheta} y\big)\,\Big\rangle_{x\,y}\Big\rangle_\vartheta,
\label{fpe.qtau}\\
\chi&=& \Big\langle\Big\langle g^\prime (\overline u_\vartheta + \sigma_{u_\vartheta} x)\Big\rangle_x\Big\rangle_\vartheta\,
\label{fpe.chi}\,,\\
\hat {\cal C}(0) &=& \int_{-\infty}^{+\infty} {\rm d}\tau\, [q(\tau) - q]\,,
\label{fpe.hC0}
\end{eqnarray}
In Eq.~(\ref{fpe.qtau}) the average $\langle \ldots \rangle_{x\,y}$ is over {\em correlated\/} normal random variables $x,y\sim{\cal N}(0,1)$ with correlation coefficient given by
\begin{equation}
\rho_{u_\vartheta}(\tau) \,= \frac{{\cal C}_{u_\vartheta}(\tau)}{{\cal C}_{u_\vartheta}(0)}=\, \frac{\sigma^2 {\rm e}^{-\kappa |\tau|} + J^2 \hat {\cal C}(0)}{\sigma^2 + J^2 \hat {\cal C}(0)}\,.
\label{fpe.rho}
\end{equation}

The $u_0$-dependent order parameters of our system are now given by the solution of Eqs.~(\ref{fpe.m})-(\ref{fpe.rho}) supplemented by the self-consistency equation (\ref{ubar}) defining the  $\overline u_\vartheta$. An analytical characterization of the fixed-points is not readily available and we have to resort to numerical analysis.

\subsection{\label{sec:AnalysisFPEs} Analysis of the Self-Consistency Equations}
In this section, we present our analysis of the fixed point equations describing the stationary dynamics of the system. In particular, we will be taking the error function,
\begin{equation}
g(x)={\rm erf}(x)=\frac{2}{\sqrt{\pi}}\int_0^x{\rm d}y\,{\rm e}^{-y^2}
\label{erf-feedback}
\end{equation}
as the sigmoid function that governs the non-linear feedback in the dynamics. This choice of feedback function has the advantage that it simplifies some of the Gaussian averages needed in the evaluation of Eqs.~(\ref{fpe.m})-(\ref{fpe.hC0}) To fully exploit this feature, we further assume that $I\sim{\cal N}(I_0, \sigma_I^2)$, so that one can combine the two Gaussians $z$ and $I$ in Eq.~(\ref{ubar}) into one. Likewise, we keep $\sigma$ constant across the ensemble of effective single node problems. These choices allow for some simplifications in evaluating the averages appearing in the original fixed point equations. E.g., evaluating $m_\vartheta$ gives
\begin{equation}
m_\vartheta=\langle{\rm erf}(\overline{u}_\vartheta+\sigma_{u_\vartheta} x)\rangle_x={\rm erf}\left(\frac{\overline{u}_\vartheta}
{\sqrt{1+2\sigma_{u_\vartheta}^2}}\right)\ ,
\end{equation}
with now  
\begin{equation}
\overline{u}_\vartheta=\frac{1}{\kappa}\Big[J_0m+I_0+\sqrt{\sigma_I^2+J^2q}\,z+\alpha J^2\chi m_\vartheta + 
\sigma_0u_0\Big]\, .
\label{ubarn}
\end{equation}
The same simplifications can be made for the other order parameters, allowing us to rewrite the set of fixed point equations as
\begin{eqnarray}
m &=& \Bigg\langle{\rm erf}\left(\frac{\overline u_\vartheta}{\sqrt{1 + 2\sigma_{u_\vartheta}^2}}\right)\Bigg\rangle_\vartheta,
\label{fpe.m new}\\
 q(\tau) &=& \Bigg\langle\Bigg\langle {\rm erf}\Big(\overline u_\vartheta +\sigma_{u_\vartheta} x\Big)\nonumber\\
& & ~~~~~~\times{\rm erf}\left(\frac{\overline u_\vartheta + \rho_{u_\vartheta}(\tau) \sigma_{u_\vartheta} x}{\sqrt{1+2(1-\rho_{u_\vartheta}^2(\tau))\sigma_{u_\vartheta}^2}}\right)\Bigg\rangle_{x}\Bigg\rangle_\vartheta\,,
\label{fpe.qtau new}\\
\chi&=& \frac{1}{\sqrt{\sigma_I^2+J^2q}}\Bigg\langle\left( z\,{\rm erf} \left(\frac{\overline u_\vartheta}{\sqrt{1 + 2\sigma_{u_\vartheta}^2}}\right)\right)\Bigg\rangle_\vartheta\,
\label{fpe.chi new}\,,\\
\hat {\cal C}(0) &=& \int_{-\infty}^{+\infty} {\rm d}\tau\, [q(\tau) - q]\,,
\label{fpe.c-hat new}
\end{eqnarray}
where, given our current system specifications, the average $\langle\dots\rangle_\vartheta$ now corresponds to an average over the Gaussian $z$ and the $\kappa$ distribution. 

To further accelerate the numerics we follow \cite{KuehnBoes93} and avoid solving the self-consistency problem Eq.~(\ref{ubarn}) for $\overline u_\vartheta$ for every member of the $\vartheta$ ensemble, by using monotonicity of the self-consistent solution $\overline u_\vartheta=\overline u_\vartheta(z)$ of Eq.~(\ref{ubarn}) to replace the $z$ average by $\overline u_\vartheta$ integrations instead. To do so we require the Jacobian of the transformation, which from the $z$-derivative of Eq.~(\ref{ubarn}), one obtains as
\begin{equation}
\frac{{\rm d}z}{{\rm d}\overline u_\vartheta}=\frac{1}{\sqrt{\sigma_I^2+J^2q}}\left[\kappa-\frac{2\alpha J^2\chi\,\exp\left(-\frac{\overline u_\vartheta^2}{1+2\sigma_{u_\vartheta}^2}\right)}
{\sqrt{\pi(1+2\sigma^2_{u_\vartheta})}}\right]\ .
\label{eq:Jac}
\end{equation} 
Following the reasoning in \cite{KuehnBoes93} we realise that for large values of $\alpha J^2\chi/\kappa$ the $\overline u_\vartheta$ distribution will have a gap corresponding to a jump in the self consistent solution of $m_\vartheta$. The critical condition for a jump in the distribution is given by
 \begin{equation}
 \frac{2\alpha J^2\chi}{\kappa\sqrt{\pi(1+2\sigma^2_{u_\vartheta})}}=1
\end{equation}
with $\overline u_\vartheta(z)$ jumping from the negative to the positive solutions of
\begin{equation}
\overline u_\vartheta=\frac{1}{\kappa}\alpha J^2\chi{\rm erf}\left(\frac{\overline u_\vartheta}{\sqrt{1+2\sigma_{u_\vartheta}^2}} \right)\, .
\end{equation}
This concludes the analysis of the general theoretical framework. We now turn to results.
\section{\label{sec: Results} Results}

In order to structure our presentation of results, it is useful to recall that --- in the absence of symmetry breaking
fields, i.e. for $I_i +\sigma_0 u_0 \equiv 0$ --- the system described by Eq.~(\ref{eq: langevin}) has a global $\mathbb{Z}_2$ symmetry $u_i \leftrightarrow - u_i$. Due to the presence of interactions, this symmetry can be spontaneously broken, giving rise to ferro-magnetic or spin-glass like phases \cite{KuBovH91, ShiiFuk92} at sufficiently low noise levels (and for sufficiently small values of the $\kappa_i$). If couplings are symmetric and if their ferro-magnetic bias is sufficiently small, the system may in fact exhibit exponentially (in system size) many meta-stable states in the zero noise limit \cite{FukShii90, Waugh+90}. Recent work \cite{FyodKhor16} has in fact demonstrated that a large number of stationary states of the noiseless dynamics continues to exist in a broad class of non-linearly interacting systems when constraints such as symmetries of interactions are dropped.

In the absence of symmetry breaking fields, phases with spontaneously broken symmetries are usually separated by sharp phase-boundaries from phases where these symmetries remain unbroken. In the context of modelling the evolution of interacting prices, however, a situation without any symmetry breaking fields in the evolution equations (\ref{eq: langevin}) would have to be regarded as {\em highly atypical}. Transitions, if any, between phases of broken and unbroken symmetries would therefore typically appear to be rounded if described in terms of the order parameters $m$, $q$ and $\chi$ appearing in the theory. It would therefore not make too much sense to precisely locate phase-boundaries which wouldn't exist as sharp boundaries for virtually {\em any\/} realistic parameter setting. In such a situation the primary interest would be to locate regions in parameter space where we expect the existence of ferro-magnetic or spin-glass like phases. We will endeavour to do this in Sect.~\ref{sec: PS}, taking properties of the phase structure that exists in the absence of symmetry breaking fields as a guidance.

Having identified interesting regions in parameter space, we will in Sect.~\ref{sec: RD} analyse distributions of log-returns for representative parameter values within these regions of interest, evaluating them for various time scales defined relative to the time scale $\gamma^{-1}$ of the slow $u_0$ process that mimics the effect of macro-economic conditions. In Sect.~\ref{sec: CP} we explore the phenomenon of collective pricing by investigating the distribution of equilibrium (log-)prices and in particular the effect that interactions between prices have on that distribution. In Sect.~\ref{sec:MSVolC}, finally, we attempt to underpin our hypothesis concerning the relation between the existence of many long-lived states in a system and the phenomenon of volatility clustering by setting up and simulating a system for which we know - at least partially - the structure of some of its meta-stable states.

\subsection{\label{sec: PS}Phase Structure}

Here, we briefly discuss the phase structure of the model, with an eye mainly towards identifying regions in parameter space were we would expect phases with glassy properties characterized by a large number of meta-stable states. The authors of \cite{KuNeu08} went some way in that direction by analysing macroscopic properties of attractors in the noiseless ($\sigma_i\equiv 0$) limit of the dynamics. In particular it was shown that the mean reversion constants $\kappa_i$, taken to be homogeneous across the system in \cite{KuNeu08}, would play a role analogous to temperature.

Continuing on the assumption of Gaussian $I_i$ made in Sect.~\ref{sec:AnalysisFPEs}, and assuming that the $\sigma_i$ are homogeneous across the network, $\sigma_i\equiv \sigma$, we have 7 parameters characterizing the system, viz. $J_0$ and $J$ determining the mean and variance, and the parameter $\alpha$ quantifying the degree of asymmetry of the couplings, as well as the mean $I_0$ and variance $\sigma_I^2$ of the distribution of the $I_i$, the strength $\sigma$ of the noise in the dynamics, and $\kappa_0$, the mean of the $\kappa$ distributions that we will consider in this paper. Unless stated otherwise, results presented in the figures below, were in fact obtained by choosing an exponential $\kappa$-distribution for the mean reversion constants $\kappa_i$. 

There can be no question of exploring this 7-dimensional parameter space completely. Fortunately we find that collective properties of the system are in the interesting region of parameter space fairly robust against parameter changes, so we will restrict ourselves to highlighting a few of the most important trends.

\begin{figure}[h!]
\includegraphics[scale=0.6]{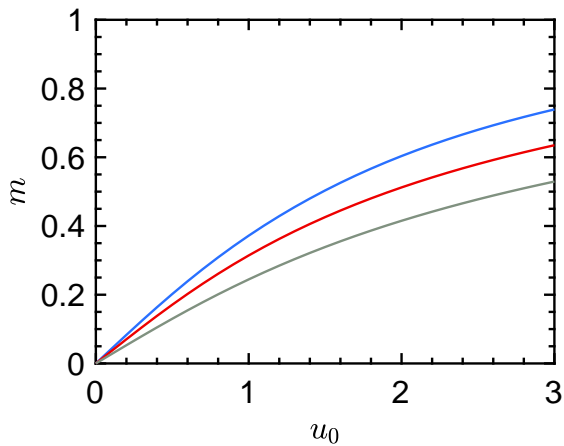}
\caption {Magnetization as a function of $u_0$, shown for three different values $\kappa_0$ of the mean of the kappa distribution. Here, we take $J_0=J=0.5$, with $\alpha=0.5$, wile $I_0=0$, $\sigma_I^2=0.1$ and $\sigma=0.1$. From top to bottom the curves correspond to $\kappa_0=0.2, 0.7$, and 1.2, respectively.}
\label{fig:mag-u0}
\end{figure}

In Fig.~\ref{fig:mag-u0} we show the behaviour of the stationary macroscopic magnetization $m$ as a function of the value $u_0$ of the slow process, using an unbiased $I_i$ distribution with $I_0=0$, and $\sigma_I^2 =0.1$.  Note that parameters characterizing the distribution of couplings and the strength $\sigma$ of the dynamic noise are chosen such that there is no spontaneous `ferro-magnetic' order in the $u_0\to 0$-limit. We also note that increasing the mean $\kappa_0$ of the $\kappa$ distribution has an effect analogous to increasing the temperature, in that it reduces the degree of macroscopic (ferromagnetic) order.

\begin{figure}[h!]
\includegraphics[scale=0.6]{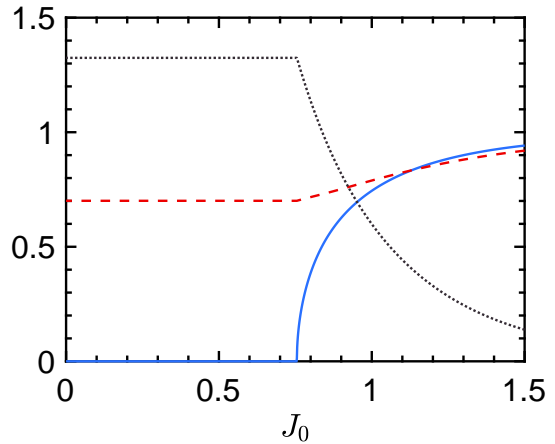}
\caption {(Color online) Magnetisation $m$ (blue full line), time-persistent correlation $q$ (red dashed line) and integrated response $\chi$ (black dotted line) as functions of $J_0$ in the absence of any global symmetry breaking fields, i.e. for $I_0=u_0=0$. Other parameters are $\sigma_I^2=0.1$, so there is a local random field, $J=0.5$, $\alpha =0.5$, $\kappa_0=0.2$, and $\sigma=0.1$. The figure shows the appearance of a ferro-magnetic phase as $J_0$ is increased beyond $J_0^c\simeq 0.75$. For $J_0 < J_0^c$ the system is in a frozen `spin-glass' like phase.}
\label{fig:J0dependence}
\end{figure}

Fig.~\ref{fig:J0dependence} illustrates that in the absence of global symmetry breaking fields, the system  exhibits a sharp second-order phase transition to ferro-magnetic order as the value of the ferro-magnetic bias $J_0$ in the coupling distributions is increased above a critical value $J_0^c$. For values of the other parameters as given, the system is in a frozen `spin-glass' like phase for $J_0 < J_0^c \simeq 0.75$. Transitions to ferro-magnetic order could also be induced by reducing the noise level $\sigma$ at sufficiently large ratios of $J_0/J$ and for sufficiently low $\kappa_0$. In a similar vein transitions into the spin-glass like phase could be induced by lowering the noise level at sufficiently small $J_0/J$-ratio, again provided $\kappa_0$ is sufficiently small. Alternatively one could chose to lower $\kappa_0$ at sufficiently small value of $\sigma$ to induce these transitions.
  
\begin{figure}[h!]
\includegraphics[scale=0.06]{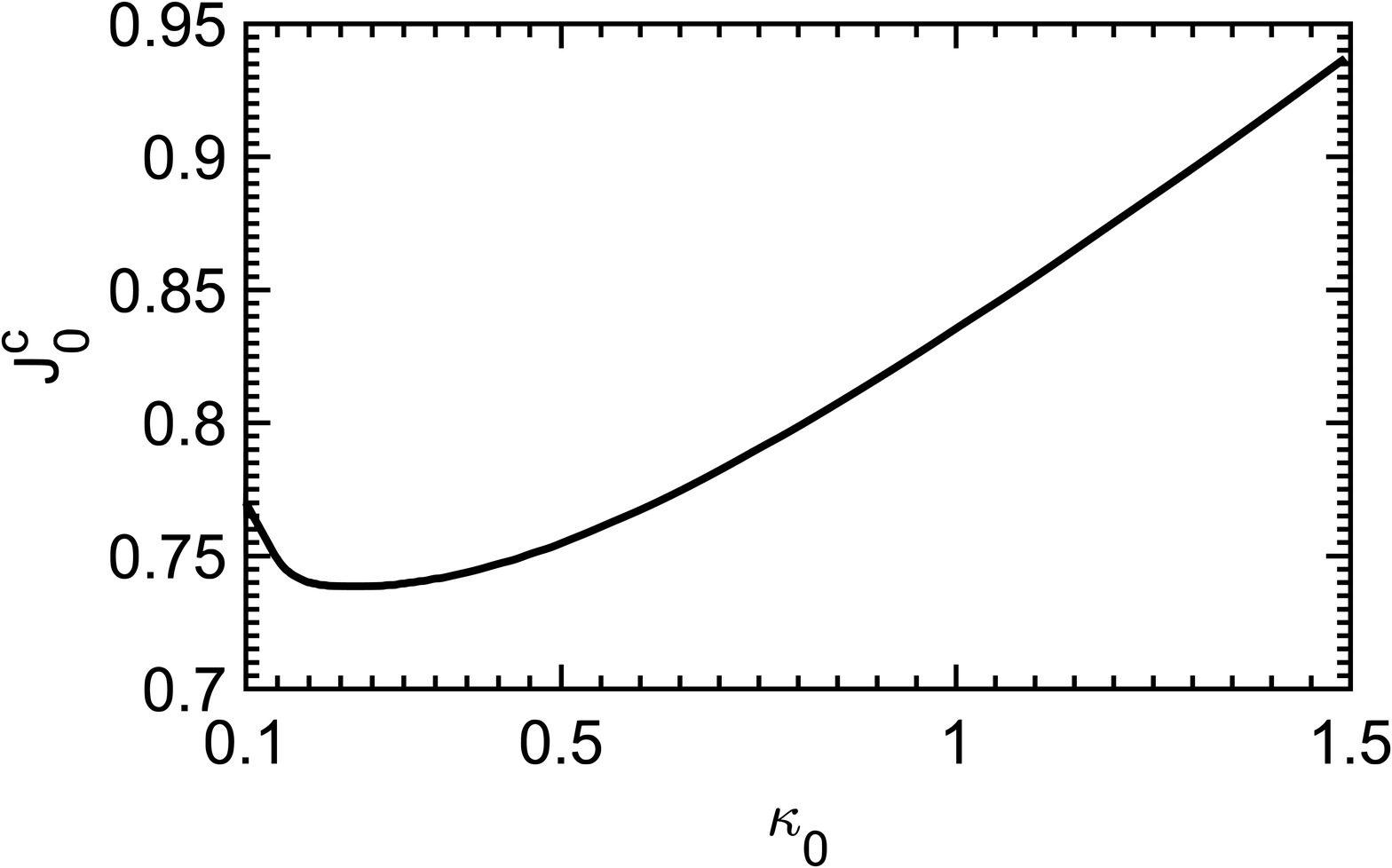}
\caption {Phase boundary separating a spin-glass like phase at small values of $J_0$ from a ferro-magnetic phase at larger values of $J_0$ as a function of $\kappa_0$ in the absence of global symmetry breaking fields, i.e. for $I_0=u_0=0$, but $\sigma_I^2=0.1$. Other parameters are $J=0.5$, $\alpha=0.5$, and $\sigma = 0.1$}
\label{fig:phaseboundary}
\end{figure}

Fig.~\ref{fig:phaseboundary} shows the phase boundary separating a spin-glass like phase at small values of $J_0$ from a ferro-magnetic phase at larger values of $J_0$ as a function of $\kappa_0$ in the absence of global symmetry breaking fields. For such a phase boundary to exist the noise level $\sigma$ has, of course, to be sufficiently low. It is expected that a spin-glass like phase will continue to exist even in the presence of weak symmetry breaking fields, in analogy to what is known for the SK model \cite{AlmeidaThouless78}.

\subsection{\label{sec: RD}Return Distributions}
We now look to compute distributions of log-returns across the ensemble of interacting assets. (In what follows, we will, somewhat loosely refer to them as return distributions). To begin with, we consider the distribution of returns for an arbitrary member of the $\vartheta$-ensemble, and so we need to consider the statistics of differences,
\begin{equation}
\Delta u_\vartheta \equiv u_\vartheta(t)-u_\vartheta(t')\ ,
\label{eq:differences}
\end{equation}
omitting time-arguments on the l.h.s. for simplicity.

We will always consider late times such that $\gamma t\gg1$ and $\gamma t'\gg1$ in order to ensure the slow process is in equilibrium. Three time scales naturally arise following the above criteria:

(i) The quasi-stationary regime for which $\gamma|t-t'|\ll 1$. In this regime we regard the fast process as stationary at a given value of the slow process. This constitutes as looking at times for which the macroscopic characterisation of the system state remains constant. 

(ii) An intermediate time scale defined by $\gamma|t-t'|={\cal O}(1)$. Explicitly, this involves looking at return distributions for stationary fast processes paramaterized by $u_0(t)$ and $u_0(t')$, for which correlations between the two slow processes still exist.

(iii) The long time scale, which we define as $\gamma|t-t'|\gg 1$, so that even the slow process has decorrelated.

In any case, we will be interested in variations induced by both the $\vartheta$ distribution and for generality, the $u_0$ statistics too. It could, however, also be of interest to inspect return distributions conditioned on specific values of the slow process.

\subsubsection{Quasi-Stationary regime}
Here, we look at the distribution of returns for the fast process in equilibrium for a given value $u_0$ of the slow process. This implies we look at time differences such that $u_\vartheta$ obeys the equation of motion Eq.~(\ref{ss-proc3}) for all times of interest. Naturally, we are interested in the limits $\kappa t,\, \kappa t'\gg 1$ which again allows us to define three distinct time scales of interest in the quasi-stationary regime itself. We will refer to these as

\begin{center}
\begin{tabular}{rll}
(i) & short: 	& $\kappa|t-t'|\ll 1$\ ,		\\

(ii) & medium:~~~& $\kappa|t-t'| = {\cal O}(1)$\ ,	\\

(iii) & long:	& $\kappa|t-t'|\gg 1$\ .
\end{tabular}
\end{center}
We note that some initial regularization of the $\kappa$ distributions (upper and lower cutoffs) may be needed to make the definition of these time windows and some of the arguments below well defined for all members of the $\vartheta$-ensemble; regularizations/cutoffs can then be removed at the end of each calculation in question. In order not to overburden the presentation, we will, however, not explicitly retrace and document these steps in what follows.

Using the solution given in Eq.~(\ref{eq:eqbm sln}), we see that 
\begin{equation}
\Delta u_\vartheta =\int_0^t{\rm e}^{-\kappa (t-s)}\eta(s)\,{\rm d}s-\int_0^{t'}{\rm e}^{-\kappa (t'-s')}\eta(s')\,{\rm d}s'\
\end{equation}

As $\eta$ is a Gaussian noise, we find that the returns for a single member of the ensemble of effective single site processes in the quasi-stationary regime are normally distributed, i.e. 
\begin{equation}
\Delta u_\vartheta \sim {\cal N}\left(0, \frac{\sigma^2}{\kappa}\left(1-{\rm e}^{-\kappa|t-t'|}\right)\right). 
\label{Delta-u-normal}
\end{equation}

For the short time scale defined above, one may expand the exponential appearing in the variance, entailing that the $\kappa$ dependence vanishes (at first order in the expansion),
\begin{equation}
\Delta u_\vartheta \sim {\cal N}\left(0, \sigma^2|t-t'|\right)\ .
\end{equation}
At these very short time separations the return distribution thus exhibits simple diffusive broadening. If $\sigma$ is taken to be constant across the ensemble, this result remains true across any portfolio of assets traded in the market.

At the long and intermediate time scales within the quasi-stationary regime there will of course be a $\kappa$-dependence of individual returns. However, if we concern ourselves with the distribution $p(\Delta u)$ of returns across the ensemble of processes, we can obtain it  by averaging the above over the $\vartheta$ distribution,
\begin{equation}
p(\Delta u)=\int{\rm d}\vartheta P(\vartheta)p(\Delta u_\vartheta)
\label{av-over-theta}
\end{equation}

\begin{figure}[h!]
\includegraphics[scale=0.65]{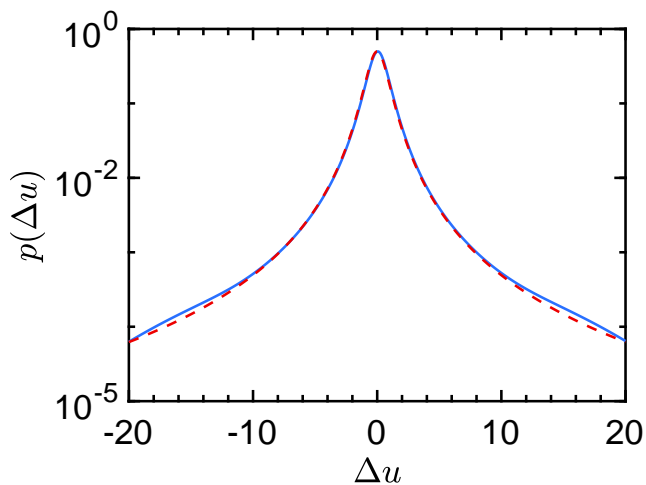}
\caption {(Colour online) Return distribution in the quasi-stationary regime evaluated for $\kappa_0|t-t'|=20$ (red dashed line) compared with the analytic  prediction for its asymptotic behaviour Eq.~(\ref{eq:asymptotic}) (blue full line) for an exponential $\kappa$ distribution with $\nu=1$.}
\label{fig:short RD}
\end{figure}

In general, this integral has to be done numerically. A simplification is possible for very large time separations within the quasi-stationary regime, for which the exponential correction in the variance in Eq.~(\ref{Delta-u-normal}) can be neglected, and $\Delta u_\vartheta \sim {\cal N}\left(0, \frac{\sigma^2}{\kappa}\right)$. Keeping $\sigma$ constant across the ensemble, the only $\vartheta$ component to average over in this limit then is $\kappa$. An analytically closed form for the distribution of returns across the ensemble can then be obtained if we assume the $\kappa$ to be $\Gamma$- distributed,
\begin{equation}
P(\kappa)=\frac{1}{\kappa_0 \Gamma(\nu)}\, \Big(\frac{\kappa}{\kappa_0}\Big)^{\nu-1}\exp (-\kappa/\kappa_0)\ .
\label{Gamma}
\end{equation}
Here $\kappa_0$ is a scale parameter which also defines the mean of the $\kappa$ distribution, while $\nu$ determines its actual shape. For this family of $\kappa$ distributions we obtain
\begin{equation}
p(\Delta u)=\frac{\sqrt{\kappa_0}}{\sqrt{2\pi \sigma^2}} \frac{\Gamma(\nu+\frac{1}{2})}{\Gamma(\nu)}\left(1+\frac{\kappa_0 (\Delta u)^2}{2\sigma^2}\right)^{-(\nu +1/2)}\ .
\label{eq:asymptotic}
\end{equation} 
Within this family of return distribution we observe power law  tail behaviour, $p(\Delta u) \sim (\Delta u)^{-\mu}$ for $|\Delta u| \gg 1$, with $\mu= 1+2 \nu$. We note that the case $\nu=1$ would correspond to an exponential $\kappa$ distribution, which would be the distribution naturally selected by the maximum entropy principle for a strictly positive random variable with a prescribed mean, in which case the tail exponent would be $\mu=3$. In Fig.~\ref{fig:short RD} we compare this analytical asymptotic result with that of a full numerical evaluation of the distribution of returns across the ensemble for $\kappa_0|t-t'| =20$, observing excellent agreement between the two already for moderate time separations.

Although we are unable to evaluate the full distribution of returns across the $\vartheta$-ensemble for intermediate time-separations, a quantity that we can evaluate in closed form for all time-separations in the quasi-stationary regime is its variance $\langle (\Delta u)^2\rangle = \langle \langle (\Delta u_\vartheta)^2\rangle \rangle_\vartheta$, in which the inner average is the variance of the return distribution for a given member of the $\vartheta$-ensemble, specified in Eq.~(\ref{Delta-u-normal}), and the outer average is over the $\vartheta$-distribution. With specifications as before, i.e. keeping $\sigma$ constant across the ensemble, the only $\vartheta$ component to average over is once more the $\kappa$-distribution. For $\Gamma$-distributed $\kappa$ as specified above we obtain
\begin{equation}
\langle (\Delta u)^2\rangle = \frac{\sigma^2}{\kappa_0 (\nu-1)} \Bigg[1 - \frac{1}{(1+\kappa_0 |t-t'|)^{\nu-1}}\Bigg]
\end{equation} 
for $\nu\ne 1$. In the $\nu \to 1$ limit this specializes to
\begin{equation}
\langle (\Delta u)^2\rangle\big|_{\nu=1} = \frac{\sigma^2}{\kappa_0} \log\big(1+\kappa_0 |t-t'|\big)
\end{equation} 
In the limit of very short time separations, this reproduces a diffusive broadening $\langle (\Delta u)^2\rangle \sim \sigma^2 |t-t'|$ of the variance which is independent of properties of the $\kappa$-distribution, as observed earlier.

It is worth pointing out that the return distributions in the quasi-stationary regime are independent of the global $u_0$ process, and in fact independent also of other parameters characterizing the interactions, as the $u_0$-dependent means $\overline u_\vartheta$, which do depend on the interaction parameters, cancel when taking differences. 

\subsubsection{Intermediate and Long Time Scales}
We now turn our attention to the case where the fast process is in equilibrium at two different values for $u_0$.  In particular, for a single member of the ensemble evaluation of Eq.~(\ref{eq:differences}) gives
\begin{eqnarray}
\nonumber\Delta u_\vartheta &=& \overline u_\vartheta(t)-\overline u_\vartheta(t')+\int_0^t{\rm e}^{-\kappa (t-s)}\eta(s)\,{\rm d}s\\
& &-\int_0^{t'}{\rm e}^{-\kappa (t'-s')}\eta(s')\,{\rm d}s'\
\end{eqnarray}

In contrast to the quasi-stationary regime, the time-dependent mean values $\overline u_\vartheta$ determined by the  values of $u_0$ at two different times $t$ and $t'$ now explicitly appear in the returns. We also expect that on this timescale, the fast noise processes have decorrelated with one another. Therefore, the distribution of returns for a given member of the ensemble and for given values of $u_0(t)$ and $u_0(t')$ now normal with non-zero mean and given by
\begin{equation}
\Delta u_\vartheta\big|_{u_0(t),u_0(t')} \sim {\cal N}\left(\Delta\overline u_\vartheta(t,t'), \sigma^2_{u_\vartheta}(t)+\sigma^2_{u_\vartheta}(t')\right)\ ,
\end{equation}
where 
\begin{eqnarray}
\Delta\overline u_\vartheta(t,t') &=& \overline u_\vartheta(t)-\overline u_\vartheta(t')\nonumber\\
\nonumber&=& \frac{1}{\kappa}\Big[J_0(m_{t}-m_{t'})+\sigma_0(u_{0}(t)-u_{0}(t'))\nonumber\\
& & + \alpha J^2(\chi_{t}m_{\vartheta,t}-\chi_{t'}m_{\vartheta,t'}) \Big] \ .
\end{eqnarray}
Here we denote an order parameter $A$ of interest by $A_{t}$ to denote its value in equilibrium for a given value $u_{0}(t)$ of the slow process $u_0$ at time $t$. As before, we are interested in the return distribution across the whole ensemble, which is obtained by averaging over the $\vartheta$-distribution. In addition to this, we can either look at return distributions for a range of specific $u_0$ values, or we may choose to average over their distribution. Since the $u_0$ term mimics the state of global economic behaviour, this average corresponds to return distributions across all market conditions. As the $u_0$ statistics are Gaussian, the joint distribution becomes easy to write down allowing us to perform this average in a straightforward manner. Finally, the return distribution across the ensemble of processes, across all market conditions is written down as
\begin{eqnarray}
\nonumber p(\Delta u)&=&\nonumber \int{\rm d}\vartheta\,{\rm d}u_0(t)\,{\rm d}u_0(t')\, P(\vartheta)\,p(u_0(t),u_0(t'))\\
& & ~~~~~~~~~~\times p(\Delta u_\vartheta|u_0(t),u_0(t'))
\end{eqnarray}

Differences between the intermediate and long time scales arise in through differences in the the joint distribution $p(u_0(t),u_0(t'))$ for the slow process. In the first case, the market conditions are still correlated whilst in the long time limit these correlations no longer persist. In both cases, we find that returns across the portfolio maintain their power-law distributed tails.  

\begin{figure}[h!]
\includegraphics[scale=0.65]{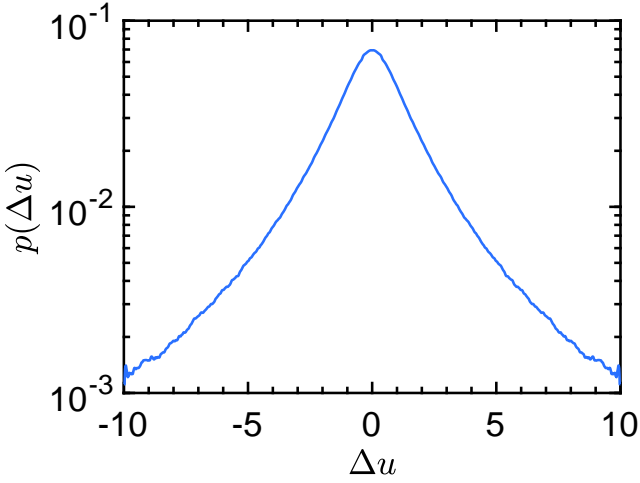}
\caption {Distributions for long time log returns across the ensemble averaged over all market conditions. Again, power law tails are observed and we expect that upon suitable normalization the distributions across timescales should scale very well as is seen in microscopic simulations.}
\end{figure}

\subsection{\label{sec: CP}Collective Pricing}   
In this section we explore the phenomenon of collective pricing mentioned at the beginning of this section. More specifically, we take a closer look at the role of $ \overline u_\vartheta$ as defined in Eq.~(\ref{ubar}); we know that this quantity takes the role of the equilibrium value of the associated asset price under given macro-economic conditions as parameterized by the value of the slow $u_0$ process. 

In order to identify the {\em collective\/} interaction-mediated properties of pricing distributions, we begin by looking at the non-interacting baseline. Without interactions in the system, the combined effect of mean reversion, drift, volatility and the value $u_0$ of the slow process describing macro-economic conditions is, according to Eq.~(\ref{ubar}), to produce a mean log-price
\begin{equation}
\overline u_\vartheta = \frac{1}{\kappa}\big[I + \sigma_0 u_0\big]
\end{equation}
that depends linearly on $I$ and on the value of the $u_0$-process. Recall that $I_i=\mu_i -\frac{1}{2}\sigma_i^2$, so $I$ includes effects of drift and volatility.

Assuming a normal distribution for $I$ as above, $I\sim {\cal N}(I_0,\sigma_I^2)$ we get a normally distributed family of
mean prices at given mean reversion,
\begin{equation}
\overline u_\kappa \sim {\cal N}\Big(\frac{I_0 + \sigma_0 u_0}{\kappa},\frac{\sigma_I^2}{\kappa^2}\Big)\ ,
\end{equation}
which, upon averaging over $\kappa$ which are $\Gamma$-distributed with $\nu > -1$ according to Eq.~(\ref{Gamma}), 
gives
\begin{eqnarray}
p(\overline u) &=& \frac{\nu \kappa_0}{\sqrt{2\pi\,\sigma_I^2}} \,
\exp\Big\{-\frac{1}{2\sigma_I^2}\Big(I_0+\sigma_0 u_0\Big)^2\Big\} \nonumber\\
& & \times \beta^{-(1+\nu)/2} \exp\Big\{\frac{\gamma^2}{4\beta}\Big\} D_{-(1+\nu)}\Big(\frac{\gamma}{\sqrt{\beta}}\Big)\ ,
\label{eq:pofubar}
\end{eqnarray}
in which $D_\nu(z)$ is a parabolic cylinder function \footnote{See I.~S.~Gradshteyn and I.~M.~Ryzhik, {\em Table of Integrals Series and Products}, Academic Press (New York, 1965), No 3.462 and Ch.~9.23.}, and
\begin{equation}
\beta =\Big(\frac{\kappa_0 \overline u}{\sigma_I}\Big)^2\ , \quad \gamma=1 -\frac{\kappa_0 \overline u}{\sigma_I^2}(I_0+\sigma_0u_0)\ .
\end{equation}
Note that $\gamma^2/\beta \to {\rm const.}$~as $|\overline u| \to \pm \infty$, so the tail-behaviour of the $\overline u$ distribution is governed by the $\beta^{-(1+\nu)/2}$ term in the above expression, giving $p(\overline u) \sim \overline u \,^{-(1+\nu)}$ for $|\overline u| \gg 1$. Conversely, the singularities which the three terms in the second line of Eq.~(\ref{eq:pofubar}) exhibit as $\overline u \to 0$  (hence $\beta\to 0$) cancel, so that $p(\overline u)$ remains finite in this limit.

Having analyzed the non-interacting case, we now return to Eq.~(\ref{ubar}), and more specifically to its version  for normally distributed $I$, Eq.~(\ref{ubarn}), to study the effects of interactions on the $ \overline u_\vartheta$. As indicated at the end of Sec \ref{sec:AnalysisFPEs} the distribution of the solution $\overline u_\vartheta$ of Eq.~(\ref{ubarn}) is obtained by transforming the normal density of $z$ 
\begin{equation}
p(\,\overline u_\vartheta) =P(z)\,\left|\frac{{\rm d} z}{{\rm d} \overline u_\vartheta}\right|
\end{equation} 
in which $P(z)=\frac{1}{\sqrt{2\pi}} {\rm e}^{-z^2/2}$, with $z=z(\overline u_\vartheta)$ obtained by solving Eq.~(\ref{ubarn}) for $z$, and the Jacobian $\frac{{\rm d} z}{{\rm d}\,\overline u_\vartheta}$ of the transformation --- for the error-function feedback (\ref{erf-feedback}) --- given by Eq.~(\ref{eq:Jac}). This allows us to obtain the distribution of equilibrium prices as induced by normal variable $z$ for a fixed $\kappa$. The distribution $p(\,\overline u)$ of equilibrium log-prices over the ensemble is obtained by averaging over the $\vartheta$ distribution as before; the average once more reduces to an average over the $\kappa$ distribution, if $\sigma$ is kept constant across the ensemble. 

\begin{figure}[h!]
\includegraphics[scale=0.6]{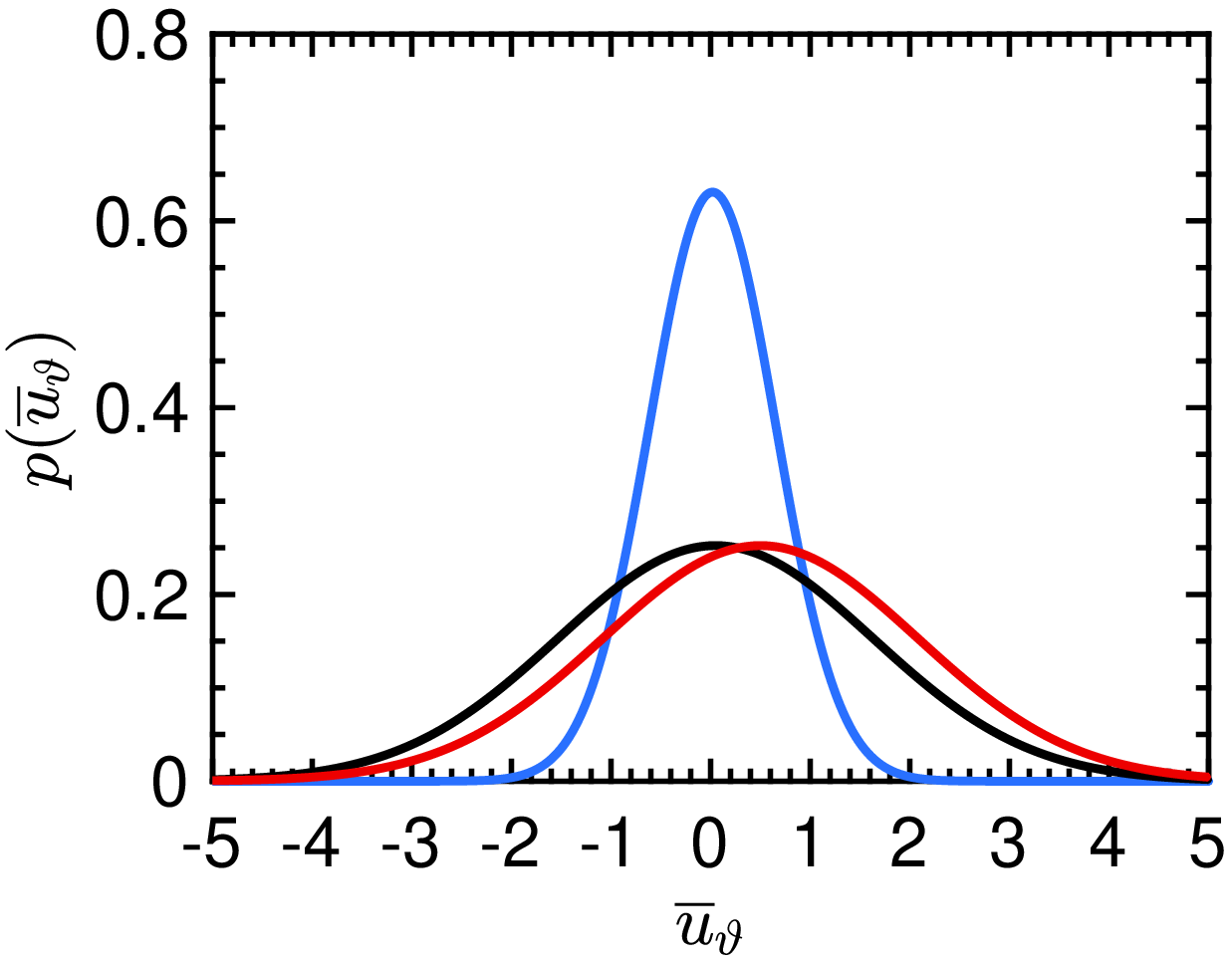}
\includegraphics[scale=0.6]{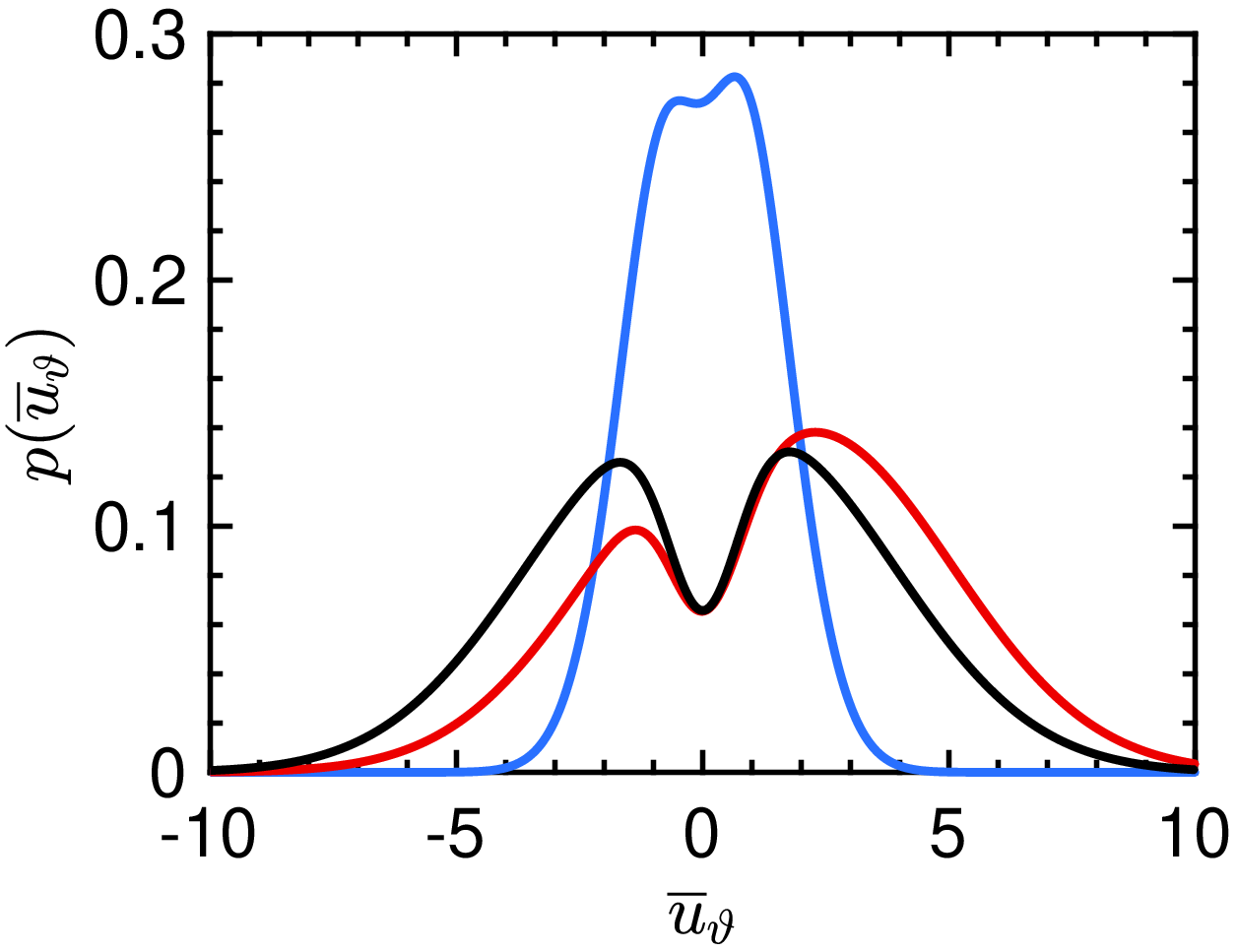}
\caption{(Colour online) Distribution of equilibrium log-prices $\overline u_\vartheta$ for the non-interacting system (first panel), compared to those of the corresponding interacting system (second panel) for selected values of the mean reversion $\kappa$ and the slow process $u_0$; note the different scales. System parameters are $I_0=0$, $\sigma_I^2=0.1$, $\kappa_0=0.2$ and $\sigma=0.1$; for the interacting system we chose $J_0=J=0.5$ and $\alpha=0.5$. For the individual curves the parameters are $(\kappa=0.5,u_0=0.1)$ (blue narrow pair distributions), $(\kappa=0.2,u_0=0.1)$ (black broad pair distributions), and $(\kappa=0.2,u_0=1)$ (red broad pair of distributions). The degree of asymmetry of the broader  distributions at smaller values of $\kappa$  increases with increasing $u_0$.}
\label{fig:pdist}
\end{figure}

We see in Fig.~\ref{fig:pdist} that interactions lead to a considerable broadening for the equilibrium distributions when compared with their non-interacting counterparts. Also the degree of asymmetry of these distributions is significantly enhanced by the interactions. A highly non-trivial effect is the systematic suppression of equilibrium log-prices which would be characterized as typical in the non-interacting system. This effect is primarily induced by the ferro-magnetic bias of the interaction, which could be induced by herding or imitation effects of agents acting in the market, or by economic fundamentals suggesting co-movement of asset prices. It  can fairly be said that this mechanism creates an interaction mediated mechanism of a market to push prices of assets to more extreme, i.e. both to very high {\em and\/} to very low values.  The effect appears to be stronger for members of the ensemble with small values of the mean reversion constant $\kappa$; it weakens for those with a larger mean reversion constant.

\begin{figure}[h!]
\includegraphics[scale=0.65]{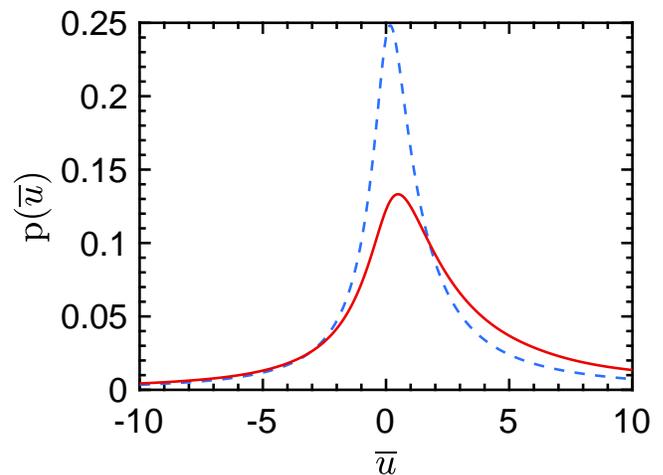}
\caption {Equilibrium distribution of log-prices across the entire ensemble for $u_0=1$, both for the interacting system (full line) and for the non-interacting system. Parameters are the same as in Fig \ref{fig:pdist}.}
\label{fig:global pricing}
\end{figure}

We may also look at the global characteristics of pricing distribution across the entire ensemble. This is achieved by averaging over the $\kappa$-distribution. For the parameter settings used, one can see in Fig.~\ref{fig:global pricing} that this smoothes out the bimodal nature observed for the sub-ensembles of assets with selected $\kappa$ values shown in Fig.~\ref{fig:pdist}, but it shows once more a considerable broadening of the distribution and a significant enhancement of the degree of asymmetry when compared with the non-interacting counterpart.

\subsection{\label{sec:MSVolC} Meta-Stable States and Volatility Clustering }   
We finally return to one of the central hypotheses underlying our modelling, namely that the complexity of real market dynamics, including in particular the phenomenon of volatility clustering could be rationalized in terms of the interplay of dynamics within (meta-stable) market states and the dynamics of occasional transitions between them. In this respect we note recent work \cite{Step+15} which identified a set of distinct market states in historical data.

The assumption behind our modelling is that market states would indeed emerge naturally as attractors of the collective (non-linear) dynamics of interacting prices. For the Gaussian couplings that we have been using in the present study, analogies with the SK spin-glass model suggest that we do in fact expect a very large number of such attractors to exist in a large region of parameter space. However, we have no way  of a-priori knowing their structure, and thus no immediate means of testing our hypothesis quantitatively.

In order to make progress in elucidating this issue we propose to look at a version of the market in which we embed a small number of {\em known\/} random attractors in the system, in order to analyze whether there is a relation between system state --- measured in terms of similarity with these known attractors --- and the observed volatility of the dynamics. For simplicity we take the system to be fully connected, and introduce couplings with a Gaussian and a Hebbian coupling component as follows,
\begin{equation}
J_{ij} = J_{ij}^{\rm (G)} + J_{ij}^{\rm (H)}\ ,
\end{equation} 
with
\begin{equation}
J_{ij}^{\rm (G)} = \frac{J_0}{N} + \frac{J}{\sqrt N} x_{ij}
\end{equation} 
and
\begin{equation}
J_{ij}^{\rm (H)} = \frac{1}{N} \sum_{\mu=1}^p \xi_i^\mu  \xi_j^\mu\ ,
\end{equation} 
in which the $\xi_i^\mu$ are i.i.d. random variable taking values $\xi_i^\mu =\pm 1$ with equal probability, and the $x_{ij}$ are normally distributed $x_{ij} \sim {\cal N}(0,1)$ and independent in pairs with $\overline{x^{\phantom x}_{ij}\, x_{ji}}=\alpha$ as in the original set-up.

\begin{figure}[t!]
\includegraphics[scale=0.6]{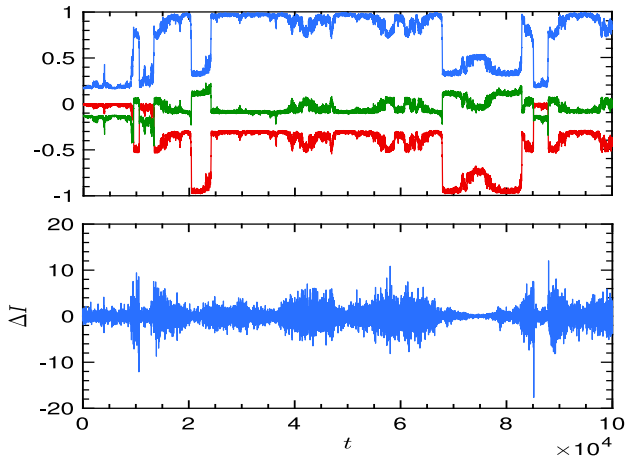}
\caption {Simulation of a market with of $N=50$ traded assets, exhibiting the relation between volatility and meta-stable state structure. The upper panel shows overlaps of the system state with three random attractors embedded in the coupling matrix in a Hebbian form as explained in the main text, while the lower panel shows returns on the index as a function of time. The other system parameters are $\kappa_0= 0.2$, $I_0=0$, $\sigma_I^2= 0.5$,  $\sigma = 0.1$, $J_0=J=0.5$, $\alpha =0.5$, and $\gamma = 10^{-4}$.}
\label{fig:overlapreturns}
\end{figure}

Figure \ref{fig:overlapreturns} presents results of a simulation of such a system of size $N=50$, with $p=3$ `patterns' embedded in the couplings, in which we simultaneously record the changes of the index, and the values of the overlaps
\begin{equation}
m_\mu(t) = \frac{1}{N} \sum_i \xi_i^\mu g(u_{it})
\end{equation} 
with the three random patterns $\{\xi_i^\mu\}$, for $\mu=1, 2$ and 3 embedded in the system. There is indeed a pronounced correlation between the volatility of the index changes and the system state as measured by the three overlaps in the system, and we believe that we can take this as a clear qualitative, and indeed semi-quantitative indication that our hypothesis of a link between meta-stable states and volatility clustering is correct for the model class under consideration.

\section{\label{sec: Summary} Summary and Discussion}

In this paper, we have provided a comprehensive analysis of the iGBM introduced in \cite{KuNeu08}. The line of reasoning leading to a model of interacting prices of this type is described in detail in that paper. Suffice it to mention here that the structure of the model follows from very generic arguments concerning the description of market mechanisms and of agents acting in a market within reduced models based on the evolution of prices alone. In the present investigation we couple the dynamics of the system to a slow Ornstein-Uhlenbeck process, which we introduce to mimic the effect of slowly evolving macro-economic conditions

We have performed a generating functional analysis of the dynamics, which maps the dynamics of the interacting system onto an ensemble of systems exhibiting a non-Markovian dynamics which is self-consistently coupled through a set of dynamic order parameters. Using a separation of time-scales argument, which assumes that the fast internal dynamics of the interacting system equilibrates at given values of the slow Ornstein-Uhlenbeck process, we are able to analyse the stationary dynamics of the system. This then allows to identify regions in parameter space where the system exhibits ferro-magnetic or spin-glass like phases.

Our analysis of the stationary dynamics (at given values of the slow driving) allows us to evaluate the distribution of log-returns for the ensemble for various time-scales, both in the quasi-stationary regime and at larger time separations. For a broad class of distributions of the mean reversion terms in the model, we find that distributions of log-returns across the ensemble are fat-tailed, exhibiting asymptotic power law behaviour broadly in line with empirical facts \cite{Gopi+99}. We note, however, that our model, as it is currently set up, does not reproduce these fat tails at the single asset level that were found empirically in \cite{Gopi+99b}. We will discuss the origin of that shortcoming, and thus possible ways to improve the model in this respect below. We are also able to evaluate the time dependent variance of the distribution of log-returns in the quasi-stationary regime, and find {\em diffusive\/} broadening in the limit of small time separations, with the broadening becoming sub-diffusive at later times. These findings are broadly in line with empirical observations.

Interestingly, our model predicts the existence of equilibrium prices for assets, and we are able to explicitly trace the influence of interactions on the distribution of equilibrium prices across the ensemble. The two main effects of collective pricing, as predicted by the iGBM are to considerably broaden the distribution of equilibrium prices in comparison with their non-interacting counterparts, as well as a significant enhancement of asymmetries characterizing such distributions for given (favourable or unfavourable) economic conditions as quantified by the value $u_0$ of the slow noise process. More specifically, we also observe a pronounced interaction-induced preference for very high or very low asset prices, which we think, deserves further study.

Note that distributions of log-returns and pricing distributions across the market are of {\em collective\/} origin, and so they can be expected to be to a certain extent independent of details of the model specifications. In particular, collective properties of the system will not depend on specific realizations of inter-asset couplings, though they may, and in general will depend of properties of coupling {\em distributions}.  This aspect could indeed provide an avenue to analysing market data within the present modelling framework which does not require to get individual couplings correct. It is also the main aspect from which the current modelling approach may eventually derive some predictive power, and might, for instance, be used to provide tools to assess market risk at a systemic level. It goes without saying that further investigations using real data will be required to get there.

One of the principal motivations for constructing the iGBM was to explore whether some of the stylized facts of financial time series could be understood in terms of effective interactions between prices of assets traded in a market, given that effective interactions between asset prices are a necessary feature of {\em any\/} model that attempts to describe market dynamics in a reduced form as a dynamics of prices alone. We have gone some way to demonstrate that this is true at the level of return distributions. Another important phenomenon is that of volatility clustering which in fact finds a quite natural explanation in terms of interacting prices. Due to the interactions, the system is expected to exhibit a large number of (dynamic and static) attractors if there is a sufficient degree of disorder and frustration. In the presence of noise, many of these attractors will survive as long-lived states, and volatility clustering is expected to arise naturally through the interplay of the dynamics within long-lived states and the dynamics of occasional transitions between them. Such transitions can occur spontaneously or be triggered by news or slowly changing macro-economic conditions. Different long lived states will be characterised by different values of their susceptibilities and so the presence of noise in the dynamics is expected to induce fluctuations with different degrees of volatility. Using simulations of a system with a partially known attractor structure, we have demonstrated in Sect.~\ref{sec:MSVolC} above that our hypothesis about a relation between meta-stable states and volatility clustering is correct at least for models of the type considered here.

In \cite{KuNeu08} the authors simulated the model using an external perturbation which they argued could represent the effect of the arrival of unexpected news (e.g. in the context quarterly reporting). The process used in that paper is difficult to implement in analytically closed form, which was one of our reasons for adopting the slow Ornstein-Uhlenbeck process, which uniformly affects all prices in a market, as a mechanism to induce transitions between meta-stable states. We believe that it is the absence of a jump-process component of the noise in the version of the model investigated in the present paper which is ultimately responsible for the fact that the model does not exhibit fat-tailed return distributions at the level of single assets. This could easily be rectified in the model formulation, by adding e.g. a Poisson jump process component to the noise, but it is likely to considerably complicate attempts at solving the model analytically. We believe it would be important to explore to what extent a model with a combination of continuous and discrete noise sources is amenable to analysis.

Our last remark refers to the presence of mean reverting forces in the iGBM, given that the existence of such forces is debated in economic circles. Within our modelling, the existence of mean reverting forces is responsible for ensuring long-term stability of the market. It would be easier to motivate the existence such forces if the $u_i(t)$ were introduced as log-prices on a {\em co-moving frame\/} as $u_i(t)\,=\,\log[S_i(t)/S_{i0} {\rm e}^{(\mu_i -\frac{1}{2} \sigma_i^2)t}]$. This modification would in the first instance eliminate the drift $I_i$ from the transformed equation (\ref{eq: after ito}), and it would suggest to introduce an iGBM formally in the same manner as was done originally, albeit with the drift term $I_i$ missing from the interacting version Eq.~(\ref{eq: langevin}) as well. Within this modified interpretation of the $u_i(t)$, the mean reversion and the interactions would have to be interpreted as mean reversion and interactions relative to an expected trend rather than relative to some fixed log-price, which might be easier to justify in economic terms. Long-term stability of the model would be saved, albeit on a co-moving frame. As an additional benefit the random symmetry breaking field $I_i$ would also disappear from the equations, which could simplify the ensuing analysis. As a downside though, such a model would likely be harder to calibrate against real market data, which should indeed be one of the next natural steps to undertake within the present project. 

{\bf Acknowledgements:} J.K.~is supported by the \mbox{EPSRC} Centre for Doctoral Training in Cross-Disciplinary Approaches to Non-Equilibrium Systems (CANES, EP/L015854/1).

\begin{appendix}
\section{Generating Functional Analysis}
\label{apx:GFA}
In this appendix we use Generating Functional Analysis (GFA) \cite{Dominicis78} to formally solve the model dynamics. We begin by introducing the generating functional in terms of source fields, $\ell$,
\begin{equation}
\label{eq:generating function}
Z[\ell|u_0] = \left\langle \exp\Big\{-{\rm i} \int {\rm d}t \sum_i \ell_i(t)n_{i}(t)\Big\} \right\rangle\ ,
\end{equation}
in which the $n_i(t) = g(u_i(t))$ are the variables in terms of which the interaction between log-prices are defined, and we condition on a realization $u_0$ of the path of the slow process representing the evolution of macro-economic conditions. The angled brackets refer to the average over all paths, which are trajectories of microscopic states. Explicitly,
\begin{equation}
Z[\ell|u_0] = \int {\cal D}{\boldsymbol{u}}\,P[\boldsymbol{u}]
\exp\Big \{-{\rm i}\int{\rm d}t \sum_{i=1}^{N}\,\ell_{i}(t)n_{i}(t)\Big\}\,,
\label{pathint}
\end{equation}
where ${\cal D}{\bm u}$ is the flat measure over a set of paths $\boldsymbol{u}\,=\,\{u_i(t)\}$, $i=1,\dots,N$  over some finite risk horizon $0\le t\le T$, and $P[\boldsymbol{u}]$ denotes the probability of these paths. The generating functional can be used to compute expectation values and correlation functions as
\begin{eqnarray}
\label{eq:generating function - compute expectation and correlation}
\langle n_i(t) \rangle &=& {\rm i}\left.\frac{\delta Z[\ell|u_0]}{\delta \ell_i(t)}\right\vert_{\ell \equiv 0}\,,\\
\langle n_j(s)\,n_i(t) \rangle &=& {\rm i}^2\left.\frac{\delta Z[\ell|u_0]}{\delta \ell_j(s)\,\delta \ell_i(t)}\right\vert_{\ell \equiv 0}\,.
\end{eqnarray}

The evaluation of the generating functional follows standard reasoning; see e.g. \cite{Dominicis78, SomZipp82, HatchettCoo04, CastCav05}. For stochastic processes described by a Langevin equation driven by Gaussian white noise, one uses $\delta$-functionals and their Fourier representations to enforce the equations of motion, which allows to transform probabilities of noise-trajectories into path probabilities. Assuming Ito-discretization for the Langevin-equation one can thus express the generating functional as 
\begin{eqnarray}
\nonumber Z[\ell|u_0] &=& \int {\cal D}\{{\bm u},\hat {\bm u}\}\exp\left\{-\int{\rm d}t \sum_i \Bigg[\frac{\sigma_i^2}{2}\hat u_i(t)^2 \right.\\ 
\nonumber & & \left. + {\rm i}\hat u_i(t)\Big(\dot u_i(t) +\kappa_i u_i(t)-I_i -\sum_j J_{ij} n_j(t)\right. \\ 
& &\left.   -\sigma_0 u_0(t)\Big) -{\rm i} \ell_i(t) n_i(t)\Bigg]\right\}\, .
\label{pathint2}
\end{eqnarray}
We are interested in evaluating the generating functional for a typical realization of disorder. This is achieved by averaging Eq.~(\ref{pathint2}) over the bond-disorder, i.e. over the $c_{ij}$ and $x_{ij}$ in terms of which the $J_{ij}$ are expressed. This disorder average factors in pairs $(i,j)$,
\begin{eqnarray}
D=\prod_{i < j} \overline{
\exp\left\{{\rm i} \int {\rm d} t \Big(\hat u_i(t) J_{ij} n_j(t) + \hat u_j(t) 
J_{ji} n_i(t)\Big)\right\}}^{~c,x} .
\end{eqnarray}
Here, we use the overbar notation to represent an average over the disorder $c$ and $x$.

Writing the $J_{ij}$ explicitly in terms of the $c_{ij}$ and $x_{ij}$ according to Eqs.~(\ref{Jij}),(\ref{jtilde}), and performing the $c_{ij}$ average in the the limit of large $N$ and finite mean connectivity $c$ we obtain
\begin{eqnarray}
D &=& \prod_{i <j} \left\{1+
\frac{c}{N}\overline{\left[\exp\left\{
\Bigg(\frac{J_0}{c} + \frac{J}{\sqrt c} x_{ij}\Bigg) \int {\rm d} t {\rm i} \hat u_i(t) n_j(t)
\right .\right .}\right .\nonumber \\
& & \left .~~ \overline{+ \left .\left . 
\Bigg(\frac{J_0}{c} + \frac{J}{\sqrt c} x_{ji}\Bigg)  \int {\rm d} t {\rm i} \hat u_j(t) n_i(t)
\right\}-1\right]}^{~x} \right\} 
\end{eqnarray}
Using the fact that $c\gg1$, we follow e.g. \cite{AnandKuehn07}, expanding the exponential to perform the $x$ average, keeping only dominant terms in the expansion in terms of inverse powers of $c$, and then re-exponentiate to write
\begin{eqnarray}
D &\simeq& \exp\Big(N\Big[J_0  \int {\rm d} t\, k(t) m(t) \nonumber \\
   & &\hspace{-0.75cm} +\frac{J^2}{2}\int {\rm d} s {\rm d} t \Big[Q(s,t)q(s,t) + \alpha G(s,t)G(t,s)\Big]\Big]\Big)\ ,
\end{eqnarray}
where we have introduced the set of one-time and two-time order parameters
\begin{eqnarray}
\nonumber m(t) &=& \frac{1}{N}\sum_{i=1}^{N}n_i(t)\,, \\
\nonumber k(t) &=& \frac{1}{N}\sum_{i=1}^{N}{\rm i}\,\hat{u}_i(t)\,,\\
\nonumber q(s,t) &=& \frac{1}{N}\sum_{i=1}^{N}n_i(s)\,n_i(t)\,,\\
\nonumber Q(s,t) &=& \frac{1}{N}\sum_{i=1}^{N}{\rm i}\,\hat{u}_i(s)\,{\rm i}\,\hat{u}_i(t)\,,\\
\nonumber G(t,s) &=& \frac{1}{N}\sum_{i=1}^{N}{\rm i}\,\hat{u}_i(s)\,n_i(t)\,. 
\end{eqnarray}
One then enforces these definitions using Dirac $\delta$-functions identities and their Fourier representations, to transform the disorder averaged generating functional into a functional integral, which to leading order in the system size $N$ can be expressed in the following compact form,
\begin{equation}
\label{GenFComp}
\overline{Z[\ell|u_0]}\,=\,\int{\cal D}\{\dots\}\exp\left\{N\left[\Xi_1\,+\Xi_2\,+\,\Xi_3\right]\right\}\,.
\end{equation}
Here, ${\cal D}\{\dots\}$ represents the functional measure over the set of macroscopic order parameter functions and their conjugates. The functionals $\Xi_1$, $\Xi_2$ and $\Xi_3$, appearing in the exponential of Eq.~(\ref{GenFComp}), are defined as
\begin{eqnarray}
\nonumber \Xi_{1} &=& J_0\int {\rm d} t\,k(t)\,m(t) 
\label{eq: Xi 1}
\, + \frac{J^2}{2}\int {\rm d}s{\rm d} t\Big( Q(s,t)\,q(s,t)\\
& & + \alpha G(s,t)\,G(t,s)\Big)\,,\\
 \Xi_{2} &=& {\rm i}\int {\rm d}t \Big( m(t)\,\hat{m}(t)\,+\,k(t)\,\hat{k}(t)\Big) \nonumber\\
 & & + {\rm i}\int {\rm d}s{\rm d}t\Big( q(s,t)\,\hat{q}(s,t) \nonumber \\
 & & \hspace{1cm}+  Q(s,t)\,\hat{Q}(s,t) +G(t,s)\,\hat{G}(t,s)\Big)\,, 
\label{eq: Xi 2}\\
 \nonumber \Xi_3 &=& \frac{1}{N}\sum_i \log \int {\cal D}\{u,\hat u\}
\exp\Big(-{\cal S}_i - {\rm i} \int {\rm d} t\, \ell_i(t)\,n(t)\Big)\\
\label{eq: Xi 3}
\end{eqnarray}
Here ${\cal S}_i$ denotes the effective local dynamic action of process $i$,
\begin{eqnarray}
{\cal S}_i &=& \int {\rm d} t ~\Big[-\frac{\sigma_i^2}{2} ({\rm i} \hat u(t))^2 
+{\rm i}\hat u(t)\Big(\dot u(t) +\kappa_i u(t) - I_i \nonumber\\
& & \hspace{1cm}-\sigma_0 u_0(t)\Big)+ {\rm i}\hat m(t) n(t) +{\rm i} \hat k(t) {\rm i} \hat u(t)\Big] \nonumber\\
& & +{\rm i} \int {\rm d} s {\rm d} t ~\Big[\hat q(s,t)n(s) n(t) + 
\hat Q(s,t) {\rm i}\hat u(s){\rm i}\hat u(t) \nonumber \\
& & \hspace{1cm} + \hat G(t,s) n(t) {\rm i}\hat u(s) \Big]\ .
\label{EffLocAct}
\end{eqnarray}
It depends on $i$ only through the locally varying parameters $(I_i,\kappa_i,\sigma_i)\equiv\vartheta_i$

One now evaluates Eq.~(\ref{GenFComp}) using the saddle point technique, which requires the macroscopic order parameters of interest to satisfy the following fixed point equations:
\begin{eqnarray}
\nonumber m(t) &=& \frac{1}{N} \sum_i \langle n(t) \rangle_{(i)}\ , \\
q(s,t)&=&\frac{1}{N} \sum_i \langle n(s) n(t) \rangle_{(i)}\ ,
\label{self-consistency}\\
\nonumber G(t,s)&=&\frac{1}{N} \sum_i \langle n(t)  {\rm i}\hat u(s) \rangle_{(i)}\ ,\quad t > s\ .
\end{eqnarray}
All other order parameters are self-consistently zero due to causality. In Eq. (\ref{self-consistency}), we use $\langle\ldots \rangle_{(i)}$ to represent an average over the dynamics of effective single site processes $i$ which takes the form
\begin{equation}
\langle \dots\rangle_{(i)}= \frac{\int {\cal D}\{u,\hat u\} (\dots)\exp\Big(-{\cal S}_i \Big)}
{\int {\cal D}\{u,\hat u\} \exp\Big(-{\cal S}_i \Big)}
\end{equation}
We note that due to causality the effective single site action simplifies to
\begin{eqnarray}
\nonumber{\cal S}_i&=&\int {\rm d} t ~\Bigg[-\frac{\sigma_i^2}{2} ({\rm i} \hat u(t))^2 
+{\rm i}\hat u(t)\Big(\dot u(t) +\kappa_i u(t) - I_i\\
\nonumber& & -J_0\,m(t) -\alpha J^2\int^t {\rm d}s\, G(t,s)n(s) -\sigma_0 u_0(t)\Big)\Bigg] \\
 & & -\frac{J^2}{2}\int {\rm d}s\,{\rm d}t\,q(s,t)\,{\rm i}\hat u(s){\rm i}\hat u(t).
\label{EfSSAct}
\end{eqnarray}

By the Law of Large numbers the saddle point equations (\ref{self-consistency}) for the order parameters can be written as averages over the distribution of the locally varying parameters $\vartheta\equiv (I,\kappa,\sigma)$,
\begin{equation*}
\frac{1}{N}\sum\limits_i\langle\dots\rangle_{(i)}\rightarrow\langle\langle\dots\rangle\rangle_\vartheta ,
\end{equation*}
as the large system limit $N\to \infty$ is taken. Here inner averages correspond to those over the dynamics of a single process with a particular parameter combination, while the outer average stands for an average over the $\vartheta$ distribution, i.e. $\langle \dots\rangle_\vartheta \equiv \int {\rm d} I\, {\rm d} \kappa\, {\rm d} \sigma\,\,p(I,\kappa,\sigma) (\ldots)$. 

One finally notes that the appearance of a contribution in the effective single-site action (\ref{EfSSAct}) which is non-local in time and quadratic in the conjugate dynamical variables $\hat u(t)$ is a manifestation of the fact that the effective single site processes are governed by {\em coloured\/} noise, while the non-local contribution involving the response function $G(t,s)$ expresses the effect that effective single site dynamics is non-Markovian. The equation of motion for the effective single site dynamics can be inferred from the effective single site action (\ref{EfSSAct}), giving
\begin{eqnarray}
\nonumber \dot{u}_{\vartheta}(t) &=& -\,\kappa\,u_{\vartheta}(t)\,+I+\,J_0\,m(t)\,+ \sigma_0u_0(t)\\
\label{EffSSProc} & &+ \alpha\,J^2\int_{0}^{t}{\rm d}s\, G(t,\,s)\,n_{\vartheta}(s)\,+\,\phi(t)\, ,
\end{eqnarray}
where we write $u(t) = u_\vartheta(t)$ when referring to single site process with local parameters $\vartheta = (I,\kappa,\sigma)$, so $n_\vartheta(t) = g(u_\vartheta(t))$, and where the coloured noise $\phi(t)$ and the dynamical order parameters appearing in Eq.~(\ref{EffSSProc})
must satisfy the self consistency equations 
\begin{eqnarray}
\langle\phi(t)\rangle & = & 0\ , \\
\langle\phi(t)\phi(s)\rangle &=&\sigma^2\delta(t-s) + J^2q(t,s)\ , 
\end{eqnarray}
and 
\begin{eqnarray}
m(t) &=&\big\langle\langle n_\vartheta(t)\rangle\big\rangle_\vartheta\ ,\\
q(t,s) &=& \big\langle\langle n_\vartheta(t)n_\vartheta (s) \rangle\big\rangle_\vartheta\ ,\\ 
G(t,s) &=& \bigg\langle\frac{\delta\langle n_\vartheta(t)\rangle}{\delta h(s)}\bigg\rangle_\vartheta\  ,\quad t> s\ .
\end{eqnarray}
We have thus reduced the original system to one comprising of an ensemble of effective single site processes characterised by the $\vartheta$-distribution with coloured noise and memory which are self-consistently determined in terms of dynamical order parameters. 

\end{appendix}
\bibliography{iGBM-2.bbl}
\end{document}